\documentclass[9pt,twocolumn,twoside]{osajnl}

\journal{ao} 

\setboolean{shortarticle}{false}
\usepackage{physics}

\usepackage{caption}

\DeclareCaptionFormat{myformat}{#1#2#3\hrulefill}
\usepackage{flushend}

\title{Simulations of mode-selective photonic lanterns for efficient coupling of starlight into the single-mode regime}

\author[1,*]{Momen Diab}
\author[1]{Aashana Tripathi}
\author[1]{John Davenport}
\author[1]{Aline N. Dinkelaker}
\author[1]{Kalaga Madhav}
\author[1]{Martin M. Roth}

\affil[1]{innoFSPEC, Leibniz Institute for Astrophysics Potsdam, An der Sternwarte, D-14482 Potsdam, Germany}

\affil[*]{Corresponding author: mdiab@aip.de}

\doi{\url{http://doi.org/10.1364/AO.421799}}

\begin{abstract}
 In ground-based astronomy, starlight distorted by the atmosphere couples poorly into single-mode waveguides but a correction by adaptive optics, even if only partial, can boost coupling into the few-mode regime allowing the use of photonic lanterns to convert into multiple single-mode beams. Corrected wavefronts result in focal patterns that couple mostly with the circularly symmetric waveguide modes. A mode-selective photonic lantern is hence proposed to convert the multimode light into a subset of the single-mode waveguides of the standard photonic lantern, thereby reducing the required number of outputs. We ran simulations to show that only two out of the six waveguides of a $1\times 6$ photonic lantern carry $>95\%$ of the coupled light to the outputs at $D/r_0 < 10$ if the wavefront is partially corrected and the photonic lantern is made mode-selective.  
\end{abstract}

\setboolean{displaycopyright}{true}

\begin{document}

\maketitle

\section{Introduction}
Although starlight arrives at the top of Earth's atmosphere with planar wavefronts that would form Airy patterns when focused by unobscured circular apertures, atmospheric turbulence distorts the wavefronts before they are collected by ground-based telescopes. Such distortion introduce random information into the wavefront which translates into the point spread function (PSF) breaking up into a speckle pattern that couples poorly with single-mode waveguides. Photonic lanterns can be used to couple atmospherically-distorted starlight into single-mode integrated optics and fibers, where the multimode speckle pattern at the focal plane is converted into multiple single-mode beams. This conversion is however only lossless if the degrees of freedom are conserved, i.e. the number of single-mode channels is at least equal to the number of supported modes at the multimode input~\cite{PL}. Since the modal content of the seeing-limited PSF increases as the telescope aperture grows or as seeing worsens, hundreds, if not thousands, of modes are required to efficiently couple all the starlight into the multimode port of the lantern which results in the signal getting split among an equal number of single-mode channels. To minimize the size of the lantern, adaptive optics (AO) may be used to first correct the received wavefront and hence reduce the modal content of the PSF to the point where only $\sim10$s of modes are required to efficiently couple the PSF of a ground-based large telescope into a multiplexed photonic device~\cite{diab_starlight_2021}.  

\begin{figure}[htbp]
    \centering
    \includegraphics[width = 1 \linewidth]{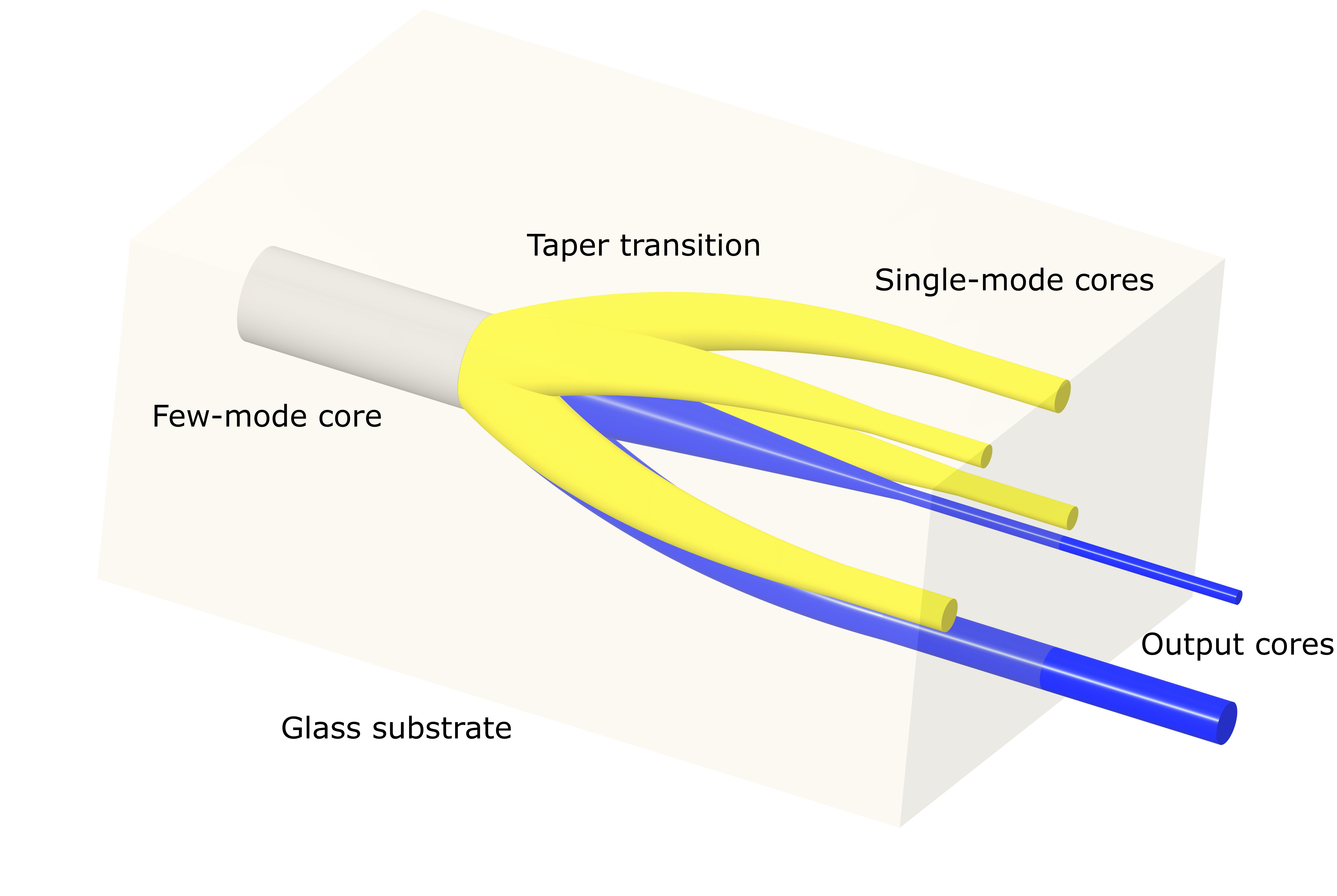}
    \caption{Layout of a $1\times 6$ MSPL inscribed in a glass substrate. Corrected starlight coupled into the few-mode core (grey) will predominantly route towards the two single-mode cores (blue) that correspond to the $LP_{0m}$ modes of the input.}
    \label{fig:bigger pic}
\end{figure}

Partially AO-corrected wavefronts result in PSFs that have a prominent core on top of a background halo~\cite{hardy_adaptive_1998}. The near symmetry of such PSFs means that they have a stronger overlap with the circularly symmetric modes of the linearly polarized (LP) modes of step-index circular fibers. By breaking the degeneracy between the single-mode waveguides, a mode-selective photonic lantern (MSPL) like the one depicted in Fig. \ref{fig:bigger pic} can be designed that converts the light coupled into a certain spatial mode to one specific output waveguide~\cite{leon-saval_mode-selective_2014}. 
For the case of partially AO-corrected PSFs, this can be exploited to transform most of the coupled multimode starlight into a subset of the total number of modes supported by the photonic lantern. Specifically, most of the light can be coupled into the waveguides associated with the circularly symmetric modes ($LP_{0m}, m = 1, 2, ...$) and thus reduce the number of single-mode channels, i.e. waveguides, that needs to be handled at the output of the photonic lantern without significant loss of light.

While conventional photonic lanterns were originally invented to accommodate fiber Bragg grating-based sky emission filters for H-band astronomy~\cite{bland-hawthorn_complex_2011}, MSPLs were first used as spatial division multiplexers (SDMs) to increase the capacity of optical communication channels~\cite{leon-saval_mode-selective_2014}. 
They have since been proposed to multiplex orbital angular momentum modes~\cite{zeng_all-fiber_2018} and to selectively amplify spatial modes in doped fibers~\cite{wittek_mode-selective_2016}. MSPLs also found applications as bending sensors~\cite{newkirk_bending_2015}, as differential group delay compensator~\cite{huang_all-fiber_2015} and made a comeback to astronomy as a way of mitigating focal ratio degradation~\cite{benoit_focal-ratio-degradation_2020}.

In this work, we present simulation results that demonstrate the potential of MSPLs as a method of reducing the number of single-mode channels for the various astronomical applications that photonic lanterns have been suggested for so far, e.g. OH suppression~\cite{bland-hawthorn_complex_2011}, reformatting~\cite{davenport_optical_2020}, multiplexed spectrographs~\cite{pimms}, and beam combiners~\cite{minardi_photonic_2012, diab_modal_2019}. First, the working principle and the design of the MSPL we considered are given. Next, we show how the modal content of starlight PSFs depends on the turbulence strength and the degree of correction. We finally present the expected performance results of using a $1\times 6$ and a $1\times 15$ MSPLs to convert AO corrected PSFs into few single-mode beams and discuss the limits beyond which this approach becomes less beneficial.      

\section{Concept and MSPL design}

Conventional photonic lanterns guide the light from a multimode core through an adiabatic transition to an array of identical single-mode cores~\cite{leon-saval_photonic_2010}. A one-to-one mapping between the spatial modes of the multimode end and the single-mode waveguides of the array can be achieved with an MSPL that has dissimilar diameters or refractive indices for the array cores. The fundamental modes supported by the single-mode waveguides will consequently have different propagation constants leading to the modes of the input port orderly coupling with the dissimilar cores one by one. Single-mode waveguides support only the fundamental mode $LP_{01}$. Higher-order circularly symmetric modes, $LP_{02}$ and $LP_{03}$, are supported by $6$ and $15$ modes waveguides, respectively. Disregarding geometry, one could directly assign different diameters or refractive indices to the cores in descending order within the allowed range. The modes from the multimode core would then occupy the single-mode cores in succession. Opting for dissimilar core sizes rather than varying the refractive indices and since the selectivity of the device can be improved by optimizing the diameters to the cores arrangement~\cite{shen_design_2018,  fontaine_geometric_2012}, we carried out an optimization of the geometry parameters 
using the beam propagation solver BeamPROP~\cite{rsoft} on CAD models of the structures. The designs assume a device written using ultrafast laser inscription (ULI)~\cite{thomson_ultrafast_2011} in a block of GLS or Eagle glass~\cite{spaleniak_integrated_2013} with refractive index contrast $\Delta = (n_{\mathit{core}} - n_{\mathit{cladding}})/n_{\mathit{core}} =  4.138\times 10^{-3}$~\cite{tepper_integrated_2017}. 

To avoid mode coupling along the transition, the adiabaticity criterion~\cite{snyder_optical_2012, yerolatsitis_adiabatically-tapered_2014}
\begin{equation}
    \left | \frac{2\pi}{\beta_1 - \beta_2} \frac{\mathrm{d}\rho}{\mathrm{d}z} \int_{A} \psi_1 \frac{\partial \psi_2}{\partial \rho} \mathrm{d}A \right | \ll 1,
    \label{eq:adiabaticity}
\end{equation}
must be fulfilled. The criterion demands that the propagation constants $\beta_1$ and $\beta_2$ of neighboring modes $\psi_1$ and $\psi_2$ that evolve slowly along the taper to be well separated if the taper length is to remain short enough for the simulations to conclude in a reasonable time. In Eq. \ref{eq:adiabaticity}, $\rho$ is the local core size and $z$ is the longitudinal coordinate, making $\mathrm{d}\rho/\mathrm{d}z$ a measure of the taper ratio, while $A$ is the structure cross-sectional area. The range of propagation constants to be filled $(\beta_{\mathit{max}} - \beta_{\mathit{min}})$ is limited by the wavelength and the normalized frequency (V-number) of the single-mode waveguides. It has an upper limit determined by the requirement for the largest waveguide in the array (corresponding to the fundamental mode of the multimode core) to remain single-mode and the smallest (corresponding to the highest order mode of the multimode core) to be $3.6$ times larger than the longest operating wavelength to have a V-number of at least $1.5$ for good field confinement of the field within the core.  The minimum separation to have the output waveguides decoupled is $\sim 30$ $\mu$m. With a device length of $50$ mm, the taper angle is $0.03$\textdegree  and therefore the taper is gradual enough to guarantee adiabaticity.

In addition to the bounds set by the propagation constants range, the maximum diameter for the single-mode cores and the minimum diameter for the multimode core depend on the operating wavelength range. The diameters are chosen such that the device will operate across the H-band ($1550$ - $1800$ nm) while keeping the number of supported modes at both ends the same.

\begin{figure*}[ht!]
    \centering
    \includegraphics[width=0.78\linewidth]{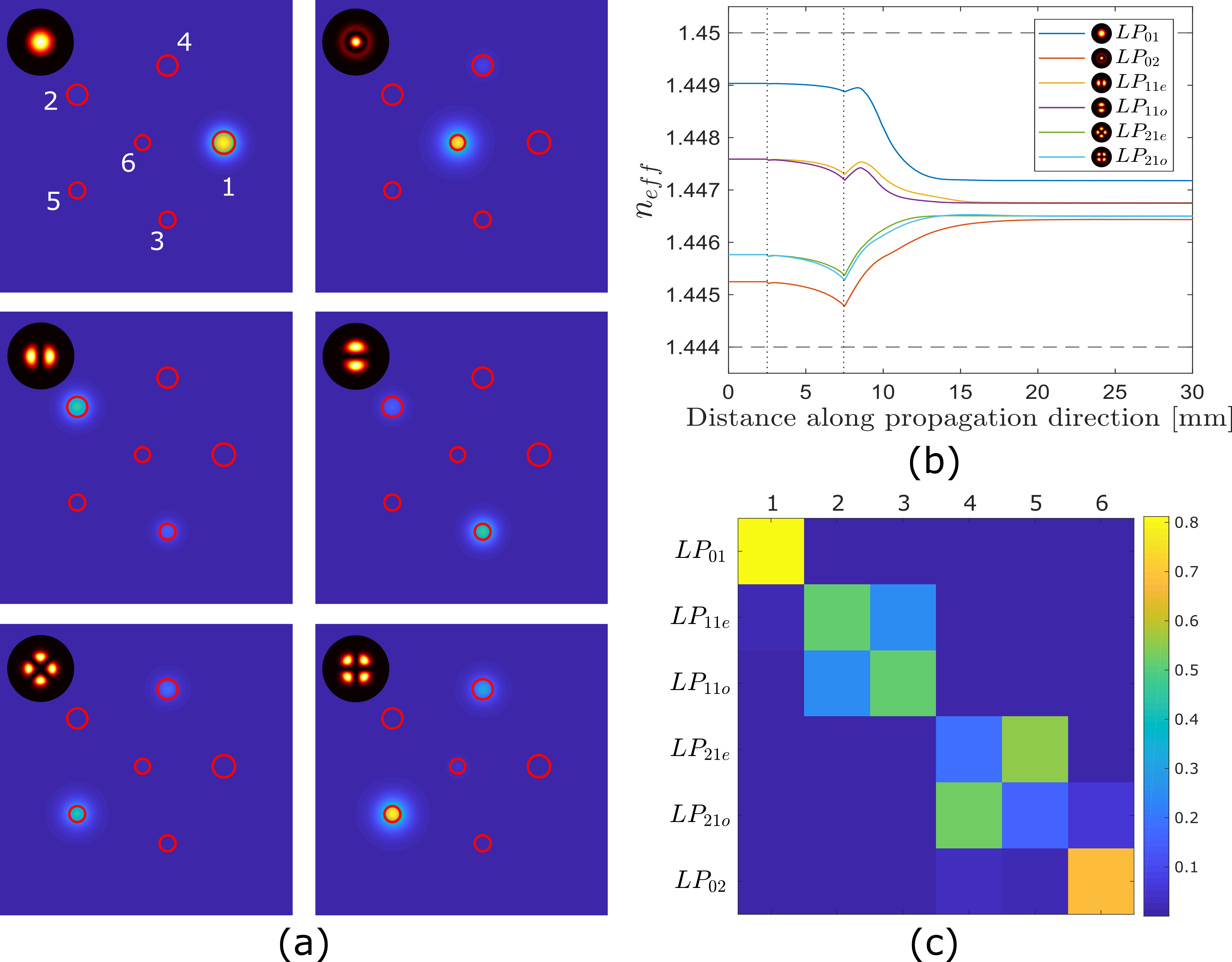}
    \caption{(a) Output patterns of the $1\times 6$ MSPL due to excitations by a pure mode. The insets show the launched mode. (b) Variation of the effective index along the propagation direction. Horizontal dashed lines indicate the cladding and the core refractive indices and the vertical dotted lines indicate the interfaces between the segments of the device. (c) Selectivity matrix illustrating the power shares of the output cores with a pure mode launched.}
    \label{fig:excitations}
\end{figure*}

Two MSPLs were designed within the constraints given above, a $1\times6$ and a $1\times15$. The anatomy of both devices consists of three segments. The input at the front facet of the glass substrate is a straight, uniform, few-mode core that can be readily spliced to a fiber. The tapered cores start at the end of the few-mode core with a matching diameter and then taper over a $5$ mm length down to the designated final diameter while fanning out from the center to form a pentagon with a central core for the $1\times6$ MSPL. The $1\times15$ MSPL has the remaining $9$ cores fanning into an outer nonagon and thus meeting the geometric requirement for lossless transition~\cite{fontaine_geometric_2012}. The last segment has cores of uniform diameters that continue to fan out at the same angle for $22$ mm to a maximum separation of $30$ $\mu$m from the center where the cores are decoupled. These segments can be identified by the jumps in the effective refractive index curves in Fig. \ref{fig:excitations}(b). 

For the $1\times 6$ MSPL (cf. Fig.~\ref{fig:bigger pic}), the multimode core has a diameter of $18.58$ $\mu$m and $\mathit{NA} = 0.13177$ and therefore supports the $6$ modes ($12$ vector modes): $LP_{01}$, $2\times LP_{11}$, $2\times LP_{21}$, and $LP_{02}$ between $\lambda = 1550$ and $1800$ $\mathrm{nm}$. The diameters for the $6$ single-mode waveguides found by the optimization are $8.5$, $7.5$, $7.5$, $5.8$, $5.8$, and $5.6$ $\mu$m, respectively. While the MSPL is highly selective for all modes, only the waveguides for $LP_{0m}$ modes are of interest here. Figure~\ref{fig:excitations}(a) shows the light spatial distribution corresponding to excitations with pure modes and Fig.~\ref{fig:excitations}(c) shows the selectivity matrix of the device, where the rows indicate how much of the total power launched into a given mode ends up at each core. 

A similar procedure is followed for the $1\times 15$ MSPL except that the diameters for the waveguides associated with the degenerate higher-order modes ($LP_{1m}, LP_{2m}, ...$) are all set to the same value to narrow the parameter space of the optimization, as only $LP_{01}$, $LP_{02}$ and $LP_{03}$ are relevant for this application.

\section{Starlight coupling into few-mode waveguides}

The atmospherically induced distortion of starlight, particularly in the field's phase, precludes efficient coupling into single-mode photonic devices, but a combination of AO and photonic lanterns can help couple light efficiently into astrophotonic instruments that allow multiplexing. Without any correction, the number of modes that the photonic lantern would need to support, $p$, (and in turn the number of channels of the device) depends on the aperture of the telescope, $D$, and the seeing condition gauged by Fried's parameter $r_0$~\cite{minardi_astrophotonics_2020}:
\begin{equation}
    p \approx \frac{\pi^2 D^2}{4 r_0^2}.
    \label{eq:modes count}
\end{equation}

A $4$ m telescope at median seeing conditions, $r_{0} = 20$ cm, in the NIR would require $\sim1000$ channels to couple the light efficiently into a multiplexed single-mode integrated optic. To fully correct such a wavefront, an AO system that has a comparable $\sim 1000$ degrees of freedom, i.e. count of the wavefront sensor (WFS) subapertures and the deformable mirror (DM) actuators, is required. However, a partial correction with only a $\sim 100$ actuators low-order system can already boost the coupling efficiency into single-mode fibers a $100$-fold and decrease the number of channels required of the photonic lanterns for full coupling to only $\sim 10$s~\cite{diab_starlight_2021}. 

Through scrambling, AO-assisted photonic lanterns can redistribute the light, more or less, equally among the single-mode channels~\cite{baudrand01} and therefore one would need to process the beams at all output ports if the flux collected by the telescope is to be fully utilized. Without scrambling, the redistribution is not equal but rather highly dependent on the time-varying environmental and atmospheric conditions, meaning again that all the channels must be used. 
An MSPL can help reduce the number of channels by routing most of the light coupled into the multimode core to only $2$ of the total $6$ channels of a $1\times 6$ MSPL. 

\begin{figure*}[ht!]
    \centering
    \includegraphics[width=0.65\linewidth]{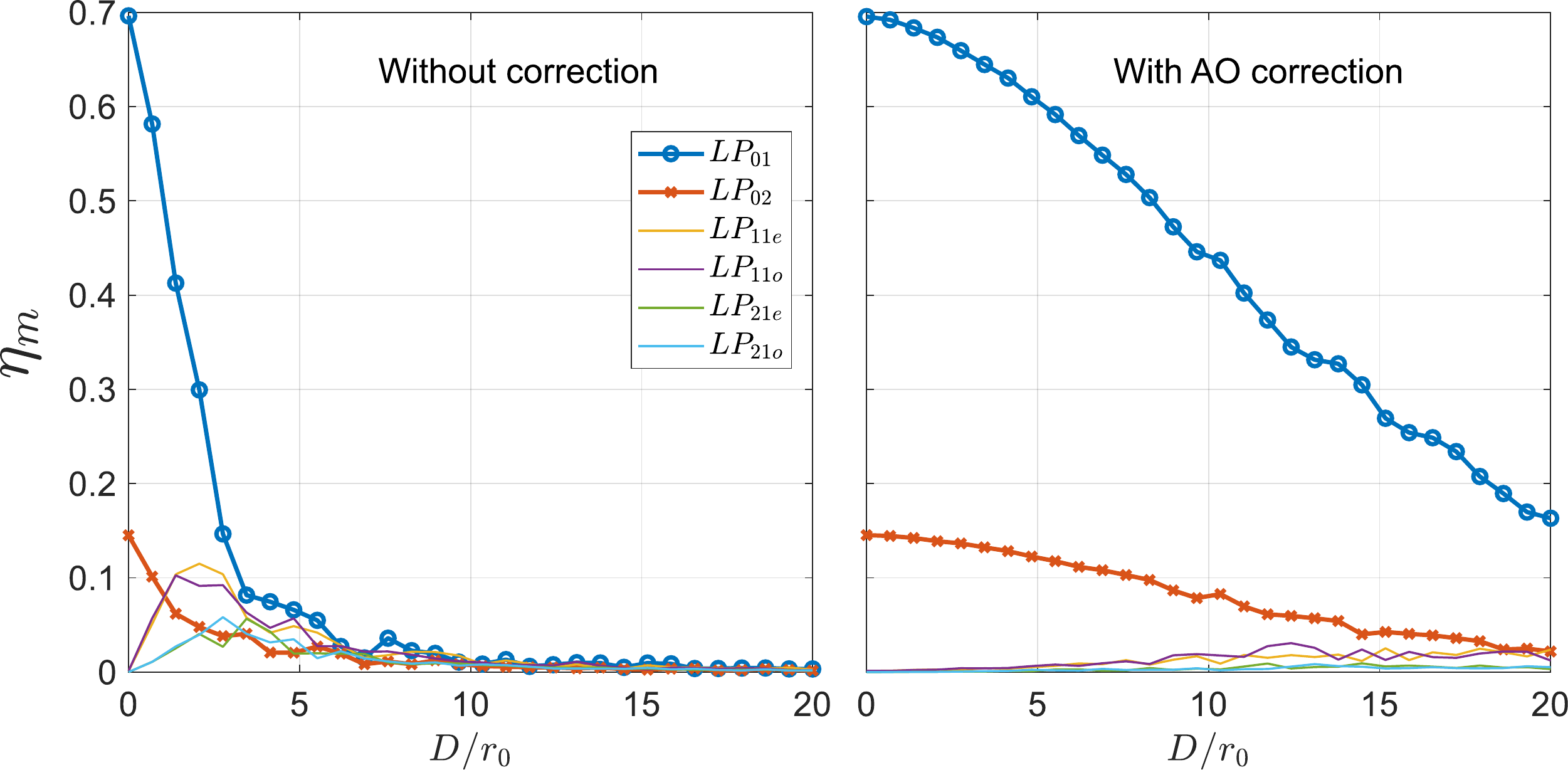}
    \caption{Contribution of the modes to coupling from an unobscured circular aperture at the optimum $f/\# = 4.83$ at $\lambda = 1550$ nm as $D/r_0$ is increased for a six-mode waveguide. Left: Without AO correction. Right: with partial AO correction.}
    \label{fig:contributions}
\end{figure*}

Figure~\ref{fig:contributions} shows the contribution of each mode to coupling into a $6$ modes waveguide as the turbulence strength $D/r_0$ is increased for both, the uncorrected and the AO-corrected cases. For this computation, $20$ seeing-limited PSFs, $\psi_E$, at each $D/r_0$ point are calculated from Kolmogorov's phase screens~\cite{byron_m._welsh_fourier_1997} and the overlap with the LP modes of a weakly-guiding, step-index circular waveguide, $\psi_i$, is evaluated to find the coupling contributions $\eta_i$ 

\begin{equation}
    \eta_i = \frac{\lvert\braket{\psi_i}{\psi_E}\rvert^2}{\braket{\psi_i}{\psi_i}\braket{\psi_E}{\psi_E}}.
	\label{eq:etai}
\end{equation}

\begin{figure*}[ht!]
    \centering
    \includegraphics[width=0.61\linewidth]{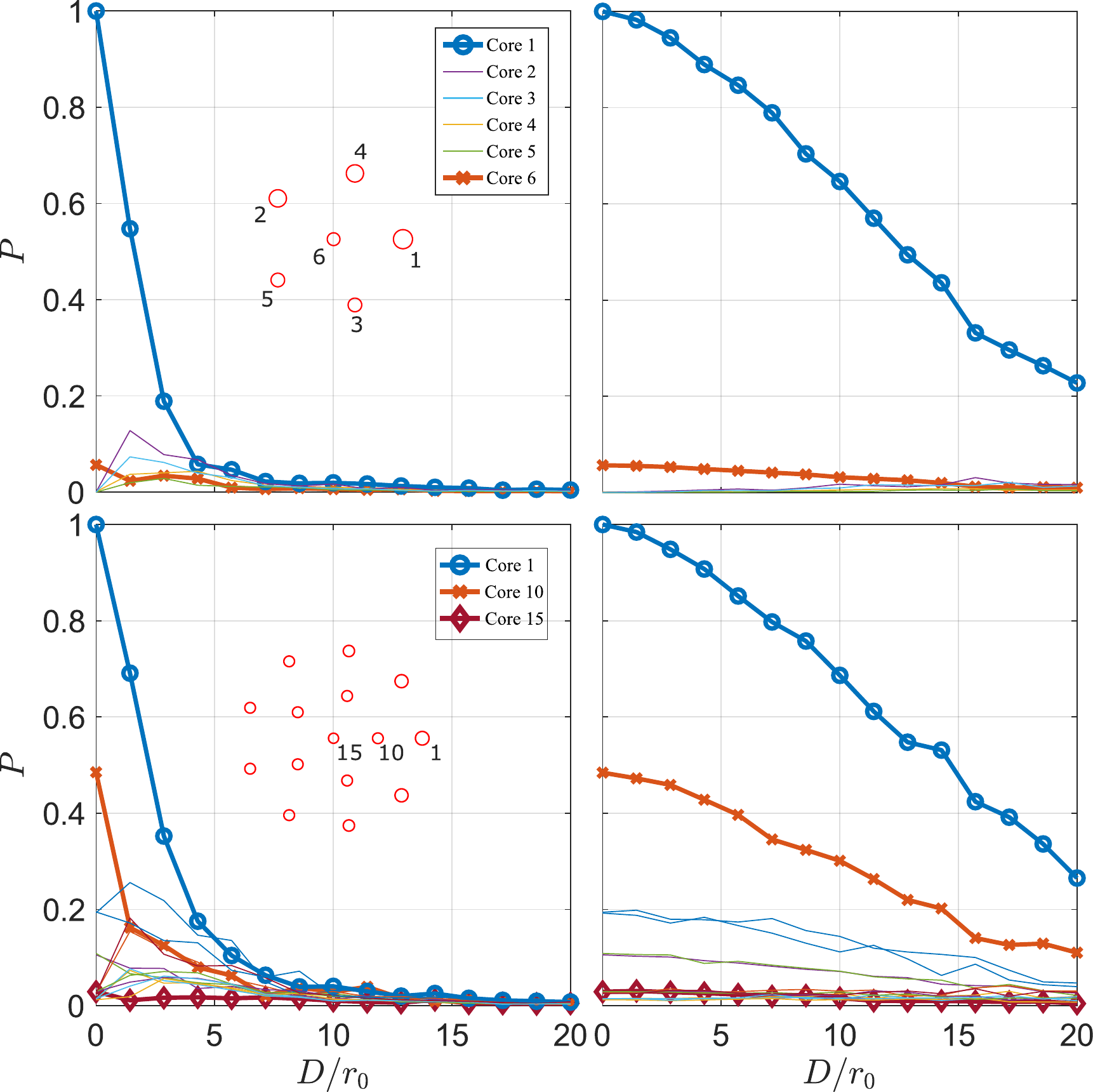}
    \caption{Normalized optical power at the single-mode output cores as turbulence strength increases for unobscured circular apertures. Insets show the cores arrangement at the output. Top: $1\times 6$ MSPL. Bottom: $1\times 15$ MSPL. Left: No correction. Right: With AO correction}
    \label{fig:6mo_noAO}
\end{figure*}

To compute the LP modes analytically, an ansatz that satisfies the symmetry boundary conditions of the cylindrical, step-index waveguide is substituted in the Helmholtz equation, $\nabla^2 \psi + n^2 k_0^2 \psi = 0$~\cite{snyder_optical_2012}. The resulting differential equations have solutions in the family of Bessel functions. Imposing the weak guidance condition, $\Delta \ll 1$, a characteristic equation is obtained that may be solved graphically to determine the parameters of the modes and subsequently their spatial distribution.

AO correction is applied to the distorted phase screens by simulating a Shack-Hartmann WFS and a DM that has $97$ actuators~\cite{gorkom_characterization_2018}. A modal reconstruction is performed to calculate the wavefront from the local slopes sensed by the WFS and find the commands for the DM~\cite{hardy_adaptive_1998}.

For the uncorrected case, the contributions of $LP_{01}$ and $LP_{02}$ are initially highest, but drop for increasing turbulence, with higher-order modes quickly contributing similar amounts. In the case where partial AO correction is applied on the distorted wavefronts, contributing the most are the $LP_{01}$ and $LP_{02}$ modes.
The same calculation was performed for the $1\times 15$ MSPL, where we still find $LP_{01}$ and $LP_{02}$ carry the largest fraction of light. $LP_{03}$, however, does not contribute.

\section{Starlight coupling into MSPLs}

To demonstrate the MSPL ability to convert the multimode starlight to a few of its single-mode outputs, focal fields of AO-corrected wavefronts at the optimum coupling $f/\#$ are computed and launched into the multimode waveguide of a model of a $1\times 6$ MSPL. The beam propagation method is then used to evolve the launch field along the MSPL and calculate the output fields at the tips of the single-mode waveguides. Figure \ref{fig:6mo_noAO} shows the dependence of the power at the outputs on $D/r_0$ for the corrected and the uncorrected cases at $\lambda = 1550$ nm. 

As the overlap of the corrected PSF with the $LP_{01}$ mode is higher than the overlap with $LP_{02}$, the $1\times 6$ MSPL will always redistribute the light unequally between the two output waveguides. Without further scrambling, this may prove problematic for certain applications, e.g. high-resolution spectroscopy, and would require a redistribution among the channels using a scrambling device for the outputs.    

\begin{figure}
    \centering
    \includegraphics[width = 1\linewidth]{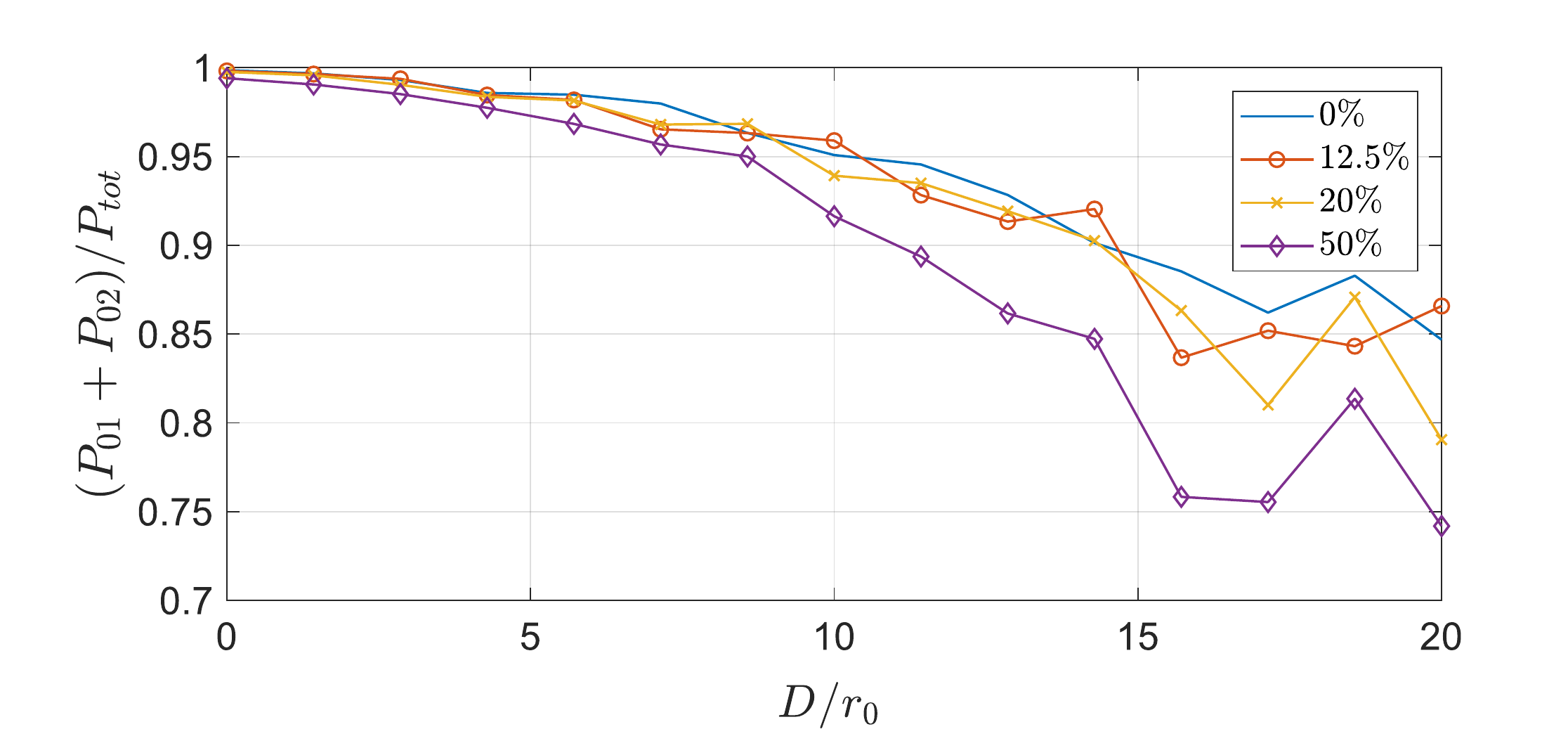}
    \caption{Share of light contained in the two $LP_{0m}$ cores for the $1\times 6$ MSPL. Curves are shown for obscuration ratio $=0\%, 12.5\%, 20\%,$ and $50\%$. }
    \label{fig:share}
\end{figure}

Of the total optical power coupled from free space into the $1\times 6$ MSPL, $>99\%$ is delivered to the two cores associated with $LP_{0m}$ at the diffraction limit. The preference for the light to steer toward those cores decreases as turbulence strength is increased but the share of the power remains $> 85\%$ at all $D/r_0 < 20$ for the unobscured telescope. Figure \ref{fig:share} shows how the share of the power in the two $LP_{0m}$ cores of a $1\times6$ MSPL depends on $D/r_o$ and the obscuration ratio of the telescope. At the diffraction limit, a central obscuration effectively redistributes part of the power from the Airy disk into the rings and thus increases the coupling into the higher-order $LP_{0m}$ modes. The total share of the power remains the same in these circularly symmetric modes as seen in Fig. \ref{fig:share}. 


\section{Conclusion and Future Outlook}

We have run simulations to demonstrate the potential of MSPLs to coupling starlight into the single-mode regime upon introducing AO correction to atmospherically distorted wavefronts. MSPLs can reduce the number of output beams that one needs to consider as compared to conventional photonic lanterns for an equal total flux delivered. For multiplexed photonic devices~\cite{watson1995}, fewer channels also mean that the signal-to-noise ratio is improved by requiring the stacking of fewer signals in post-processing after detection~\cite{diab_starlight_2021}. The MSPL is selective across the H-band where the modal count of both ports remains the same.     

Several methods can be considered to fabricate MSPLs. For the simulations presented here, we have assumed the ULI technique. With ULI, 3D structures can be written into bulk glass by moving the substrate in all three dimensions relative to a focused short-pulse laser beam. The refractive index is modified at the focus, thereby producing waveguide structures. By combining laser parameters, focusing optics, and sample movement appropriately, the position, shape, and size of each waveguide can be changed. ULI has already been used to write single-mode waveguides, directional couplers and interferometers~\cite{piacentini_space_2020}, Bragg-gratings~\cite{brown_ultrafast_2012}, photonic reformatters~\cite{pike_multi-core_2020, maclachlan_development_2016}, as well as photonic lanterns~\cite{thomson_ultrafast_2011, spaleniak_integrated_2013}. MSPLs, however, have not yet been fabricated using ULI. High positional precision and repeatability will be required to shape the individual waveguides, especially for the lowest mode field diameter (MFD) difference of 200 nm between mode-selective waveguides in our simulations. Here, the substrate motion will be crucial. For systems utilizing air-bearing translation stages, (e.g.~\cite{brown_ultrafast_2012, maclachlan_development_2016}), hardware specifications from the manufacturers state positioning repeatability of $\pm 25-100$~nm. While we expect the precision to be sufficient, experimental tests should be conducted to verify MFD and selective coupling of ULI-manufactured devices. The effect of position uncertainty might be mitigated when using the multiscan technique, where the combined index change of several displaced scans accumulates to form the final waveguide, e.g.~\cite{brown_ultrafast_2012}.

Furthermore, photonic lanterns can be produced by tapering stacks of optical fibers. Many single-mode fibers fuse together to form the multi-mode waveguide~\cite{davenport_optical_2020}. MSPLs can be produced by this method, either by using optical fibers of varying sizes~\cite{yerolatsitis_adiabatically-tapered_2014}, with similar cladding diameters but different cores~\cite{leon-saval_mode-selective_2014}, or using a multicore fiber~\cite{benoit_focal-ratio-degradation_2020}.

With a fabricated device, the simulation results reported here can be verified on an AO testbed where $D/r_0$, the degree of correction, and the obscuration ratio can be changed~\cite{diab_testbed_2020}.
After initial proof-of-concept experiments in the lab, subsequent on-sky tests would be required. As losses in the laser-written devices can accumulate to cancel any signal advantage from the MSPL, the throughput has to be evaluated. Optimization might be required to reduce losses. However, if the throughput is too low, a similar MSPL could be fabricated from a fiber stack for testing on an astronomical telescope. Both techniques can be used complementally to find suitable MSPL configurations for different telescopes and AO systems in order to improve light coupling in photonic devices under the effect of turbulence.


\begin{backmatter}
\bmsection{Funding} German Federal Ministry of Education and Research
(BMBF) (03Z22AN11).

\bmsection{Acknowledgments} The authors thank the Photonic Instrumentation (PHI) Group at Heriot-Watt University, especially Aur\'elien Benoit, for fruitful discussions on the ULI technique. 

\bmsection{Disclosures} The authors declare no conflicts of interest.

\bmsection{Data Availability} Data underlying the results presented in this paper are not publicly available at this time but may be obtained from the authors upon reasonable request.

\end{backmatter}


\bibliography{sample}

\bibliographyfullrefs{sample}


\ifthenelse{\equal{\journalref}{aop}}{%
\section*{Author Biographies}
\begingroup
\setlength\intextsep{0pt}
\begin{minipage}[t][6.3cm][t]{1.0\textwidth} 
  \begin{wrapfigure}{L}{0.25\textwidth}
    \includegraphics[width=0.25\textwidth]{john_smith.eps}
  \end{wrapfigure}
  \noindent
  {\bfseries John Smith} received his BSc (Mathematics) in 2000 from The University of Maryland. His research interests include lasers and optics.
\end{minipage}
\begin{minipage}{1.0\textwidth}
  \begin{wrapfigure}{L}{0.25\textwidth}
    \includegraphics[width=0.25\textwidth]{alice_smith.eps}
  \end{wrapfigure}
  \noindent
  {\bfseries Alice Smith} also received her BSc (Mathematics) in 2000 from The University of Maryland. Her research interests also include lasers and optics.
\end{minipage}
\endgroup
}{}

\end{document}